\documentclass[11pt]{article}
\pdfoutput=1

\usepackage[a4paper,top=3cm,bottom=4cm]{geometry}

\usepackage{Packages}

\usepackage{Definitions}

\allowdisplaybreaks[1]

\newcommand{\OfficialTitle}{Classical Integrability of the Squashed
  Three-sphere, Warped $\mathrm{AdS}_3$ and Schroedinger Spacetime via T--Duality}

\hypersetup{pdfauthor={Domenico Orlando, Susanne Reffert and Linda I. Uruchurtu},pdftitle={\OfficialTitle}}

\author{
  \begin{minipage}{.97\linewidth}
    \vspace{1cm}
    \begin{center}
      {\small
        \textbf{Domenico Orlando}$^{1}$, \textbf{Susanne Reffert}$^{1}$ and \textbf{Linda I. Uruchurtu}$^{2}$
      }
    \end{center}
    \vspace{1cm} \hspace{2cm}\begin{minipage}{.7\linewidth}
      {\itshape \footnotesize
          \begin{itemize}
          \item[${}^1$] Institute for the Mathematics and Physics of
              the Universe, \\The University of Tokyo, Kashiwa-no-Ha
              5-1-5, \\ Kashiwa-shi, 277-8568 Chiba, Japan.
            \item[${}^2$] Theoretical Physics Group, The Blackett Laboratory\\
              Imperial College London, Prince Consort Road \\
              London, SW7 2AZ, UK
          \end{itemize}}
    \end{minipage}
    \vspace{1cm}
    \end{minipage}
}

\date{}

\title{\vspace{2cm}
  {\huge   \textbf{\OfficialTitle}}
}

\begin{document}

\setstretch{1.1}

\numberwithin{equation}{section}

\begin{titlepage}
  \maketitle
  \thispagestyle{empty}

  \vspace{-15cm}
  \begin{flushright}
    IPMU10-0195 \\
    Imperial/TP/2010/LIU/02
  \end{flushright}

  \vspace{15cm}

  \begin{center}
    \textsc{Abstract}
  \end{center}
  We discuss the integrability of $2d$ non-linear sigma models with
  target space being the squashed three-sphere, warped anti--de Sitter
  space and the Schroedinger spacetime. These models can be obtained
  via T--duality from integrable models. We construct an infinite
  family of non-local conserved charges from the T--dual Lax currents,
  enhancing the symmetry of warped anti--de Sitter space and the
  Schroedinger spacetime to $\widehat{\mathfrak{sl}}_2 (\setR) \oplus
  \widehat{\mathfrak{sl}}_2 (\setR)$.
\end{titlepage}

\newpage

\section{Introduction}
\label{sec:introduction}

Our understanding of superconformal gauge theories is far from
complete. However, for specific cases, integrability has enabled us to
expand our insights and provide useful ways for determining different
properties of the underlying theory (e.g. anomalous dimensions of
operators, determination of other observables, etc.). Indeed, for the
case of $\AdS_{5}/\CFT_{4}$, the integrable structure of planar
$\mathcal{N}=4$ super Yang--Mills theory has allowed the determination
of the spectrum via tools such as the Bethe Ansatz, finite gap
methods, S--matrices, etc.~\cite{Metsaev:1998it, Bena:2003wd, Kazakov:2004qf, Beisert:2005fw, Beisert:2005bm, Beisert:2006ez, Arutyunov:2009ga}. Furthermore, integrability was also shown to be present in other \AdS/\CFT systems
such as $\AdS_{4}/\CFT_{3}$~\cite{Arutyunov:2008if, Stefanski:2008ik, Sorokin:2010wn} and $\AdS_{3}/\CFT_{2}$~\cite{Chen:2005uj, Adam:2007ws, Babichenko:2009dk}, supporting the idea that more examples of integrable systems should exist.

More recently, new examples of holographic systems have been proposed 
after novel target space metrics have been shown to emerge naturally as solutions of various three dimensional gravity theories. 
In particular, backgrounds containing \emph{squashed geometries} (spheres and anti-de
Sitter spaces) have become an interesting arena for further explorations, as such configurations have long been known in the
context of deformed \CFT{}s and black holes in string theory~\cite{Rooman:1998xf, Israel:2003ry, Israel:2004vv, Israel:2004cd, Detournay:2005fz, Orlando:2006cc}. For specific setups, it has been possible to
determine the central charge of the dual \CFT \footnote{It might be that these "central charges" are really a result of a generalised version of Cardy's formula, and the jury is still out with respect to these dual theories being conformal.} by looking at its
asymptotic symmetry algebras~\cite{ Compere:2007in, Compere:2008cv, Compere:2008cw, Anninos:2008fx, Gupta:2010ib}, but little more
is known about its precise properties and only recently, integrability
has come into play~\cite{Ricci:2007eq, Babichenko:2009dk}. Another
interesting and related example is that of the \emph{Schroedinger
  spacetime}~\cite{Son:2008ye,Balasubramanian:2008dm,Maldacena:2008wh}. This
background is invariant under the Schroedinger group which includes
translations, rotations, Galilean boosts and non-relativistic scale
transformations. Its holographic properties have been widely studied
owing to its inherent appeal to condensed matter physics, where
systems described by strongly coupled non-relativistic QFTs are common
fare~\cite{Guica:2010sw,Costa:2010cn}.

Independently, two-dimensional non-linear sigma models which are
integrable are interesting in their own right, as it has been a
longstanding question how to identify integrable
systems. In~\cite{Ricci:2007eq, Mohammedi:2008vd,Curtright:1994be} it
was shown that the T--duals of integrable systems often turn out to be
new integrable models, by explicit determination of their Lax pairs
and the construction of the infinite set of conserved charges. This
was shown in~\cite{Mohammedi:2008vd} for the \textsc{pcm} and SU(2)
sigma models. This was also discussed
in~\cite{Balog:1993es,Evans:1994hi}. Based on these results, the
authors of~\cite{Ricci:2007eq} discussed how integrability translates
from the original model to the T--dual one, focusing on the discussion
of the (bosonic) $\AdS_5\times S^5$ case which was known to be
integrable and whose T--dual model is again $\AdS_5\times S^5$. The
emerging picture was later generalized to the full superstring action
in~\cite{Beisert:2008iq}.

In~\cite{Babichenko:2009dk} the integrability approach was extended to
$\AdS_3 \times X_{7}$ backgrounds supported by \textsc{rr} fluxes in
which standard worldsheet methods cannot be applied. It was shown that
backgrounds with sixteen supercharges of the form $\AdS_3 \times S^3
\times M_4$ with $M_4=T^4$ or $S^3 \times S^1$ can be described using
a Green--Schwarz action admitting a $\mathbb{Z}_4$
grading~\cite{Berkovits:1999zq}, which in turn implies
integrability. This is due to the fact that the equations of motion
and the Maurer--Cartan equations of any ${Z}_4$ coset can be
re-expressed as flatness conditions for a Lax
connection~\cite{Bena:2003wd}.

Given that $\AdS_3\times S^3 \times M_4$ is equipped with an
integrable structure, it is natural to ask whether new examples of
integrable systems can be obtained from this background via
T--duality.
Squashed three-spheres ($\SqS^3$), warped anti--de~Sitter spaces
($\WAdS_3$) and the Schroedinger spacetime $\Sch_3$ are string theory
backgrounds obtained via T--duality from $\AdS_3 \times
S^3$~\cite{Orlando:2010ay}. Building on this observation we show that
T--duality relates non-linear sigma models with group manifold $G$
target space to their squashed counterparts $\SqG$. The construction
that we present here has two main advantages compared to the treatment
in~\cite{Orlando:2010ay}: it is easier because it is not based on
dimensional reduction and it is more general since it can be applied
to obtain the squashing in the compact directions of any Lie group. In
this sense, $\WAdS_3$ and $\SqS^3$ can be seen as the \emph{simplest
  examples}. The example of $\Sch_3$ is different in the sense that
the T--duality has to be performed in a \emph{non-compact} direction.

From our point of view, T--duality is a linear transformation of the
components of the conserved currents of the initial model. This means
that the integrability properties of the non-linear sigma model on $G$
are inherited by the T--dual model\footnote{Related works with
  T--duality in the context of integrable models and conformal sigma
  models can be found in~\cite{Balog:1993es, Miramontes:2004dr,
    Gomes:2004bw, Alday:2005ww, Arutyunov:2005nk,
    Bakas:1995hc,Sfetsos:1995ac,LopesCardoso:1994df}.} on $\SqG$. The initial manifold has
isometry $G \times G$ which is promoted via a Lax construction to
affine $\widehat{\mathfrak{g}} \oplus
\widehat{\mathfrak{g}}$~\cite{Schwarz:1995td,Lu:2008kb}. This affine
symmetry remains after the T--duality, but now the zero modes are not
anymore isometries of the target space. This is a crucial point: the
squashing preserves only $G \times T$ isometry (where $T \subset G$ is
the maximal torus), but the full symmetry of the T--dual model is
$\widehat{\mathfrak{g}} \oplus \widehat{\mathfrak{g}}$, whose zero
modes are the isometries supplemented by a set of currents generating
non-local charges that cannot be found via a Noether construction. In
the case of $\WAdS_3$ for example, even though the isometry is
$\mathrm{SL}_2(\setR) \times \mathrm{U}(1)$, 
we find the full $\widehat{\mathfrak{sl}}_2
(\setR) \oplus \widehat{\mathfrak{sl}}_2 (\setR)$ symmetry algebra.
The construction also works in the limit case, when the T--duality
connects the non-linear sigma model on $G$ to the one on
$G/\mathrm{U}(1)$. Also in this case we find an affine $\widehat
{\mathfrak{g}} \times \widehat {\mathfrak{g}}$ algebra of
symmetries. Since our analysis is classical, it is not surprising to
find that the affine algebra is actually a loop algebra, without
central element; it will not be the case once we add quantum
corrections. In this paper we consider backgrounds with only the
metric turned on; the analysis for the full type II solution will be
presented in a forthcoming publication~\cite{Reffert:2010fu}. The
results that we present will nevertheless remain true in the more
general setting with \textsc{rr} fields.

\bigskip 

The plan of this paper is as follows. In
Section~\ref{sec:general-framework} we describe the general
framework. We review how squashed backgrounds are obtained via
T--duality in Section~\ref{sec:squashed-groups}. In
Section~\ref{sec:integr-princ-chir} we review the integrability of the
principal chiral model and in Section~\ref{sec:integr-squash-groups}
extend the results to the T--dual squashed groups.
Section~\ref{sec:examples} is devoted to explicitly working out the
flat currents (Lax connections) and conserved charges of our
examples. The squashed three-sphere $\mathrm{Sq}S^{3}$ is discussed in
Section~\ref{sec:SqS3}, three-dimensional warped anti-de Sitter space
in Section~\ref{sec:WAdS} and the Schroedinger spacetime $\Sch_3$ in
Section~\ref{sec:example-thre-schr}.  The procedure is very similar to
the one discussed in~\cite{Alday:2005ww,
  Frolov:2005ty,Kluson:2007md}. We show that the original $G \times G$
isometry group ($\mathrm{SU}(2)\times \mathrm{SU}(2)$ for $\SqS^3$ and
$\mathrm{SL}_2(\setR) \times \mathrm{SL}_2(\setR)$ for $\WAdS_3$ and
$\Sch_3$) is realized non-locally in the T--dual model via the
appearance of non-local currents. This generalizes and extends the
results of~\cite{Kawaguchi:2010jg} where the $\mathrm{SU}(2) \times
\mathrm{U}(1)$ isometry of $\SqS^3$ was extended to a hidden Yangian
algebra, but non-local currents were not considered. Finally,
Section~\ref{sec:conclusions} is devoted to a discussion and summary
of our results, with some outlook on future work. \\
Appendix~\ref{sec:hier-vari-princ} reviews in detail the integrable
hierarchy of variations for the principal chiral model.
Appendix~\ref{sec:deform-hier-squash} discusses the infinite hierarchy
of the squashed groups after T--duality.
Appendix~\ref{sec:geod-squash-groups} details the geodesics on
squashed groups, providing further insights into their nature.

\section{General framework}
\label{sec:general-framework}

In this section, we review the general framework. In
Section~\ref{sec:squashed-groups}, we discuss how to construct
\textsc{nlsm} on squashed groups via T--duality from principal chiral
models with group manifold target space $G$, generalizing the
treatment in~\cite{Orlando:2010ay}.  In
Section~\ref{sec:integr-princ-chir}, we review the classical
integrability of the \textsc{pcm}. In
Section~\ref{sec:integr-squash-groups}, we show the classical
integrability of the \textsc{nlsm} on squashed groups obtained via
T--duality by explicitly constructing a one-parameter family of non-local
conserved currents from the T--dualized Lax currents. In order to do
so, we first have to find a gauge transformation on the original model
to bring its currents into a form which is readily T--dualizable. It
is this gauge transformation which ultimately leads to the non-local
nature of the dualized Lax currents, which cannot be obtained from a
Noether procedure.

\subsection{Squashed groups via T--duality}
\label{sec:squashed-groups}

Consider a non-linear sigma model on a target space $M$ with coordinates
$( u^i, z )$. If $\partial_z$ is a space-like \emph{Killing vector}, the metric on $M$
can be put into the form
\begin{equation}
  \label{eq:fibration-metric}
  \left( \begin{tabular}{c|c}
    $G_{ij}(u) + f_i (u) f_j(u)$ & $f_i(u)$ \\ \hline
    $f_i(u)$ & $1$ 
  \end{tabular}\right) \, .
\end{equation}
The corresponding action on a surface $\Sigma$ with signature $(-,+)$
is given by
\begin{equation}\label{eq:oriact}
  S[u^i,z] = \int_\Sigma G_{ij}(u) \di u^i \wedge \st \di u^j + \left( \di z + f_i(u) \di u^i \right) \wedge \st \left( \di z + f_j (u) \di u^j \right) \, ,
\end{equation}
where the $u^i $ and $z$ are maps
\begin{align}
  u^i : \Sigma &\to \setR \,,\\
   (x,t) &\mapsto u^i(x,t)\,.
\end{align}

We want to T--dualize on $z$ using a first-order formalism (see
\cite{Rocek:1991ps,Buscher:1987sk}).  Introduce a gauge
field $A$ and the Lagrange multiplier $\widetilde z$:
\begin{equation}\label{eq:Tact}
  S[u^i, A, \widetilde z] = \int_\Sigma
  G_{ij}(u) \di u^i \wedge \st \di u^j + \left( A + f_i(u) \di u^i \right)
  \wedge \st \left( A + f_i(u) \di u^i \right) - 2 \widetilde z\, dA\,.
\end{equation}
The equations of motion for $\widetilde z$ give:
\begin{equation}
    \di A = 0 \hspace{1em}    \Rightarrow \hspace{1em} A = \di z \, ,
\end{equation}
which leads back to the original action~(\ref{eq:oriact}). On the other hand, the
equation of motion for $A$ leads to
\begin{equation}
  \label{eq:T-duality-condition}
  \st \di \widetilde z = A + f_i(u) \di u^i = \di z + f_i(u) \di u^i\,.
\end{equation}
Note that $ \di z + f_i(u) \di u^i $ is the current associated to the
Killing vector $\partial_z$.

Plugging the condition in Eq.~\eqref{eq:T-duality-condition} back into
the action~(\ref{eq:Tact}) we get the T--dual sigma model on the space
$\wt M$ where the fibration has become trivial, \emph{i.e.} the
geometry is now a direct product, and a $B$ field appears. The action is
\begin{equation}
  S[u^i,\widetilde z] =
  \int_\Sigma G_{ij}(u) \di u^i \wedge \st \di u^j + \di \widetilde z
  \wedge \st \di \widetilde z - 2 \di \wt z \wedge f_i(u) \di u^i \, , 
\end{equation}
which corresponds to a target space metric on $\wt M$ given by
\begin{equation}
    \left( \begin{tabular}{c|c}
      $G_{ij}(u) + f_i (u) f_j(u)$ & $0$ \\ \hline
      $0$ & $1$ 
    \end{tabular}\right) \, .  
\end{equation}

From a more abstract point of view one can think of the initial target
space $M$ as the total space of an $S^1$ fibration whose fiber is
parametrized by the coordinate $z$ and whose base space $N$ has
coordinates $u^i$ and metric $G_{ij}$:
\begin{equation}
  \begin{CD}
    S^1 @>>> M  \\
    @.      @VVV\\
    {} @. N
  \end{CD} \, .
\end{equation}
After the T--duality transformation, we obtain a non-linear sigma
model with target space $\widetilde{M}$ where the fibration has been traded
for a direct product,
\begin{equation}
  \wt M = N \times S^1 \, ,
\end{equation}
and the circle is parametrized by the new coordinate $\widetilde z$.

\emph{Principal chiral models} (\textsc{pcm}) are natural examples for this construction,
since a Lie group $G$ is a principal bundle with fiber $H$ and base
$G/H$ for any closed subgroup $H$:
\begin{equation}
  \begin{CD}
    H @>>> G  \\
    @.      @VVV\\
    {} @. G/H
  \end{CD} \, .
\end{equation}  

In our main examples we will concentrate on geometries described by
fibrations in which the total space is $G \times \mathrm{U}(1) $, and
the fiber is a rational linear combination of one direction in the
Cartan subgroup of $G$ and the extra $\mathrm{U}(1)$:
\begin{equation}
  \begin{CD}
    \mathrm{U}(1) @>>> G \times \mathrm{U}(1) \\
    @.      @VVV\\
    {} @. \mathrm{Sq  G}
  \end{CD} \, ,
\end{equation}  
where the base space $\mathrm{SqG}$ is the squashed group with metric
\begin{equation}
  \label{eq:ds2-SqG}
  \di s^2 [ \mathrm{SqG} ] = \di s^2 [G] + \tanh^2 \Theta \, j_C \otimes j_C \, ,  
\end{equation}
where $\Theta$ is a real parameter related to the radius of the
$\mathrm{U}(1)$,\footnote{As discussed in~\cite{Orlando:2010ay},
  $\Theta$ can assume only a discrete set of values in a String Theory
  embedding.}  and $j_C$ is the current in the Cartan
direction\footnote{In our normalizations the metric on $G$ is decomposed as
  \begin{equation}
    \di s^2[G] = - \sum_{a=1}^{\dim G} j_a \otimes j_a \, ,
  \end{equation}
  where $j_a = \Tr[ g^{-1} \di g \, T_a]/\Tr[T_a^2]$, and $T_a$ are
  the generators of $\mathfrak{g} = \mathop{\mathrm{Lie}}(G) $.
}.  The
T--dual geometry is the Cartesian product
\begin{equation}
  \wt M = \mathrm{SqG} \times S^1 \, .
\end{equation}
In the limit case $\Theta \to \infty$ the T--dual geometry is the product
\begin{equation}
  \wt M = \left( G/ \mathrm{U}(1) \right) \times \mathrm{U}(1) \, .
\end{equation}

\begin{proof}
  To show this, decompose $g \in G \times \mathrm{U}(1)$ as
  \begin{equation}
    g = l \exp \left[ \imath \psi T_C + \imath \frac{y}{\Tr[T_C^2]} T_{r+1} \right] \, ,  
  \end{equation}
  where $T_C$ is a generator of the Cartan subalgebra of $\mathfrak{g}
  = \mathop{\mathrm{Lie}}(G)$, $T_{r+1}$ generates the extra
  $\mathfrak{u}(1)$, and $l \in G/\mathrm{U}(1)$. The metric on the
  manifold $G \times \mathrm{U}(1)$ is written as:
  \begin{equation}
    \di s^2[G \times \mathrm{U}(1)] = \frac{ \Tr[ \di g \di g^{-1}]}{\Tr[ T_C^2]} = \frac{\Tr[ \di l \di l^{-1}]}{\Tr[T_C^2]}  + \di \psi^2 + \di y^2 \frac{\Tr[T_{r+1}^2] }{\Tr[T_C^2]^3} 
   - \frac{2\imath  \di \psi \Tr[ l^{-1} \di l T_C]  }{\Tr[T_C^2]} \, .
  \end{equation}
  Now introduce the variable
  \begin{equation}
    \label{eq:change-of-variables}
    \alpha = \psi - \frac{y}{\Tr[ T_C^2]} \, .
  \end{equation}
  The group element takes the form
  \begin{equation}
    \label{eq:g-u(1)-decomposition}
    g = l \exp \left[ \imath \alpha T_C + \frac{\imath y}{\Tr[T_C^2]} \left( T_C + T_{r+1} \right) \right] = k \exp \left[  \frac{ \imath y}{\Tr[T_C^2]} \left( T_C + T_{r+1} \right) \right] \, , 
  \end{equation}
  where $k \in G$. The metric becomes
  \begin{equation}
    \di s^2[ G \times \mathrm{U}(1)] = \frac{ \Tr[ \di k \di k^{-1} ] }{\Tr[ T_C^2]}+ \frac{\di y^2}{\Tr[T_C^2]^2} \left( \frac{\Tr[T_{r+1}^2]}{\Tr[T_C^2]} + 1 \right) - \frac{2 \imath \di y}{\Tr[T_C^2]} \frac{ \Tr[ l^{-1} \di l T_C  + \imath \di \alpha T_C^2]}{\Tr[T_C^2]} \, .
  \end{equation}
  Introduce the parameter $\Theta$,
  \begin{equation}
    \sinh^2 \Theta = \frac{\Tr[T_C^2]}{\Tr[T_{r+1}^2]} \, ,
  \end{equation}
  and rescale $y$ as
  \begin{equation}
    \frac{y}{\Tr[T_C^2]} = z \, \tanh \Theta \, ,
  \end{equation}
  then the metric becomes
  \begin{equation}
    \label{eq:GxU1-metric}
    \begin{split}
      \di s^2[ G \times \mathrm{U}(1) ] &= \di s^2 [G] + \di z^2 - 2
      \imath \tanh \Theta \di z j_C \\
      &= \di s^2[G] + \tanh^2 \Theta \, j_C \otimes j_C + \left( \di z
        - \imath \tanh \Theta \, j_C \right)^2 \, ,
    \end{split}
  \end{equation}
  where
  \begin{align}
    \di s^2 [G] &= \frac{\Tr[\di k \di k^{-1}]}{\Tr[T_C^2]} \, , & j_C
    = \frac{\Tr[ k^{-1} \di k T_C]}{\Tr[T_C^2]} \, .
  \end{align}
  The structure is precisely the same as for the metric in
  Eq.~\eqref{eq:fibration-metric}, hence by applying the T--duality
  transformation on $z$ we get the condition
  \begin{equation}
    \st \di \wt z = \di z - \imath \tanh \Theta \, j_C  \, ,
  \end{equation}
  and the corresponding T--dual metric
  \begin{equation}
    \di s^2 = \di s^2[G] + \tanh^2 \Theta \,
    j_C \otimes j_C + \di \wt z^2 \, ,
  \end{equation}
  which is the metric of a squashed group times $S^1$.
\end{proof}

The T--duality breaks the initial $G \times G \times \mathrm{U}(1)$
isometry to $G \times T \times \mathrm{U}(1)$, where $T \subset G$ is
the maximal torus. Further insights on the geometry of squashed group
manifolds can be obtained by studying their geodesics. This is done
in Appendix~\ref{sec:geod-squash-groups}.

\subsection{Integrability of the principal chiral model}
\label{sec:integr-princ-chir}

Consider the two-dimensional \textsc{pcm} for a Lie group $G$. The action is given by
\begin{equation}
  S = - \frac{1}{2} \int_\Sigma \Tr [ \di g(x,t) \wedge \st \di g^{-1}(x,t) ] \, ,
\end{equation}
where $g$ is a map from the worldsheet $\Sigma$ with coordinates
$(x,t)$ to the group $G$. The equations of motion take the form
\begin{equation}
\label{eq:PCM-EOM}
  \di\st (g^{-1} \di g) =   \di \st ( \di g\, g^{-1} ) = 0 \, .  
\end{equation}
From these equations we can read off the two conserved currents which
result from the explicit invariance of the action under the
transformations $\delta g = \epsilon\, g $ and $\delta g = g\, \overline
\epsilon$, where $\epsilon, \overline \epsilon \in \mathfrak{g} $:
\begin{align}
  j &= g^{-1} \di g \, , & \overline j &= - \di g\, g^{-1} \, .
\end{align}
Those currents are \emph{flat} and thus fulfill the
\emph{Maurer--Cartan} (\textsc{mc}) equations:
\begin{align}
  \di j + j \wedge j &= 0 \, , & \di \overline j + \overline j \wedge \overline j &= 0 \, .
\end{align}
These flatness conditions are the underlying reason for the
integrability of the model.  Note that this theory is different from
the \textsc{wzw} model which has an extra Wess--Zumino
term. Nevertheless also in this case it is possible to find an affine
algebra of symmetries, as we show in the following.

Introduce the one-parameter families of currents
\begin{align}
  J(x, t; \zeta)&= - \frac{\zeta}{ 1 - \zeta^2 } \left( \zeta \, j(x,t)
    + \st j(x,t) \right)  \, , &  \overline J(x, t; \zeta)&= -
  \frac{\zeta}{ 1 - \zeta^2 } \left( \zeta \, \overline j(x,t)
    + \st \overline j(x,t) \right)  \, , 
\end{align}
where $\zeta \in \setC$ is the \emph{spectral parameter}.
Imposing the flatness of $J$ and $\overline J$ produces two equations for
the components $(J_x, J_t)$ and $(\overline J_x, \overline J_t)$, the so-called \emph{Lax equations}:
\begin{align}
  \del_t J_x - \del_x J_t + \comm{J_t}{J_x} &= 0 \, , &
  \del_t \overline J_x - \del_x \overline J_t + \comm{\overline J_t}{ \overline J_x} &= 0  \, .
\end{align}
Each couple $(J_x, J_t)$ and $(\overline J_x, \overline J_t)$ forms a \emph{Lax pair}~\cite{Lax:1968fm}.
In order to simplify the notation, we introduce the operator
\begin{equation}
  \Lambda(\zeta) = \frac{\zeta}{1-\zeta^2} \left( \zeta + {\st} \right) \, , 
\end{equation}
and we concentrate on the left current
\begin{equation}
  J(x,t;\zeta) = - \Lambda (\zeta) j(x,t) \, . 
\end{equation}
We can expand $\Lambda(\zeta)$ in a power series as follows:
\begin{equation}
  \Lambda(\zeta) = {\zeta \st} + \zeta^2 + {\zeta^3 \st} + \zeta^4 + \dots =
  \sum_{n=1}^\infty {\zeta^{2n - 1} \st} + \zeta^{2n} \, .
\end{equation}
Observe in particular that $J(\zeta)$ has no zero mode in the
expansion in powers of $\zeta$. This is reflected in the fact that the
model admits Noether charges.
It is also useful to remark that
\begin{equation}
  \label{eq:star-Lambda}
  \st \Lambda(\zeta)\,j = \zeta \left( \Lambda(\zeta) + 1 \right)j \, . 
\end{equation}
The flatness of $J$ and $\overline J$ implies both the \textsc{eom} and
the \textsc{mc} equations. Conversely, imposing the \textsc{eom} and
\textsc{mc} equations results in the flatness of the currents. This
can be easily verified by observing that
\begin{equation}
  \di J(\zeta) + J(\zeta) \wedge J(\zeta) = \frac{\zeta}{\zeta^2 - 1} \left( \di \st j + \zeta \left( \di j + j \wedge j \right) \right) \, .
\end{equation}
The flatness of $J$ and $\overline J$ can be used to construct two infinite
sets of conserved charges. %
Introduce a \emph{Wilson line} $W(x, t; \zeta)$ such
that
\begin{equation}
  J(x, t; \zeta) = W^{-1} (x, t; \zeta) \di W(x, t; \zeta) \, .  
\end{equation}
More precisely, $W$ depends on two points $(x,t)$ and $(x_0, t_0)$ on $\Sigma$ and can be written as
\begin{equation}\label{eq:w}
  W(x, t| x_0, t_0; \zeta) = \mathop{\mathrm{P}}\left\{ \exp \left[ \int_{\mathcal{C}:(x_0,t_0) \to (x,t)} J(\xi, \tau; \zeta) \right]\right\} \, ,  
\end{equation}
where $\mathop{\mathrm{P}}$ denotes the path-ordering and
$\mathcal{C}$ is a path from $(x_0,t_0)$ to $(x,t)$. We can now define a
one-parameter family of conserved charges:
\begin{equation}
  Q(t; \zeta) = W(\infty, t | -\infty, t; \zeta) =
  \mathop{\mathrm{P}}\left\{ \exp \left[ \int_{-\infty}^\infty J_x(x,
      t; \zeta) \di x \right] \right\}\, .
\end{equation}
Using the Lax equations one can show that if the current $J$ vanishes
at spatial infinity ($J(\pm \infty, t; \zeta ) = 0$), the
one-parameter charge $Q(t; \zeta)$ is conserved%
:
\begin{equation}\label{eq:cond}
  \frac{\di}{\di t} Q(t;\zeta ) = 0 \, .
\end{equation}
Expanding in a power series in $\zeta$,
\begin{equation}
  \label{eq:conserved-charges}
  Q(t; \zeta ) = 1 + \sum_{n=0}^\infty  \zeta^{n+1} Q^{(n)}(t) \, . 
\end{equation}
The condition in Eq.(\ref{eq:cond}) is equivalent to the conservation
of the infinite set of charges (see~\cite{Luscher:1977rq}),
\begin{equation}
  \frac{\di}{\di t} Q^{(n)}(t ) = 0, \hspace{2em} \forall \,n = 0, 1, \dots  \, .    
\end{equation}
In other words, the model is \emph{classically integrable}. 
The charge $Q^{(0)}$ is written explicitly as
\begin{equation}
  Q^{(0)} = \int_{-\infty}^\infty j_t (x,t) \di x \, ,  
\end{equation}
and it is the Noether charge corresponding to the manifest symmetry $g
\mapsto g+ \epsilon g$, for $\epsilon \in \mathfrak{g}$. All the
other charges are non--local (\emph{i.e.} they cannot be written as
integrals of densities), but can always be understood in terms of
transformations $g \mapsto g + \delta g$ that leave the
\textsc{eom}~\eqref{eq:PCM-EOM} invariant (but are not invariants of
the action).  The Poisson brackets of the set of $Q^{(n)}$ form a
Kac--Moody algebra $\widehat{\mathfrak{g}}$.  This is explained in
detail in Appendix~\ref{sec:hier-vari-princ}. An alternative
description of the charges can be obtained in terms of Yangian
symmetry; this is discussed in detail in~\cite{Abdalla:548604}.

\subsection{Integrability for squashed groups from T--duality}
\label{sec:integr-squash-groups}

As remarked above, principal chiral models are natural examples for
the T--duality construction of
Section~\ref{sec:squashed-groups}. Since we wish to study their
classical integrability properties, we can think of the duality as of
a linear transformation of the current components $(J(\zeta), \overline
J(\zeta)) \mapsto (\wt J(\zeta), \wt {\overline J}(\zeta))$ that leaves the
(on-shell) flatness conditions invariant:
\begin{align}
  \di \wt J + \wt J \wedge \wt J &= 0\,, & \di \wt {\overline J} + \wt {\overline J}
  \wedge \wt {\overline J} &= 0 \, ,
\end{align}
where the T--dual Lax currents $\wt J(\zeta)$ and $\wt {\overline J}(\zeta)$
are obtained by imposing the condition in Eq.~\eqref{eq:T-duality-condition},
\begin{equation}
  \st \di \wt z = \di z + f_i(u) \di u^i \, .  
\end{equation}
In this case, in Eq.~(\ref{eq:GxU1-metric}) we have found that the
metric on $G \times \mathrm{U}(1)$ can be written as
\begin{equation}
  \di s^2[ G \times \mathrm{U}(1)] = \di s^2[G] + \tanh^2 \Theta \, j_C \otimes j_C + \left( \di z
    - \imath \tanh \Theta \, j_C \right)^2 \, , 
\end{equation}
and we want to perform a T--duality on $z$ imposing the condition
\begin{equation}
  \st \di \wt z = \di z - \imath \tanh \Theta \, j_C \, . 
\end{equation}

An important point is that in general the current $\di z - \imath
\tanh \Theta \, j_C$ does \emph{not} commute with all the currents
$j$.  This is reflected by the fact that some of the components of $J$
depend explicitly on $z$ and not only on the differential $\di z$. For
this reason it is necessary to introduce a suitable group-valued
function $h : \Sigma \to G$ and perform a gauge transformation:
\begin{equation}
  J' = h^{-1} J\, h + h^{-1} \di h \, ,   
\end{equation}
so that the new family of flat currents $J'(\zeta)$ does not depend
explicitly on $z$ and can be dualized to $\wt J'(\zeta)$. The price to
pay is that $\wt J'$ contains a \emph{zero mode} in the development in powers
of $\zeta$:
\begin{equation}
  \wt J'(\zeta) = \left. h^{-1} \di h - \Lambda(\zeta) h^{-1} j h \right|_{ \di z = \st \di \wt z - f_i(u) \di u^i } = \wt J'^{(0)} - \Lambda(\zeta) \wt j \, .
\end{equation}
Moreover, the zero mode depends explicitly on $h$ which is a non-local
object when expressed in terms of $j$.
\begin{proof}
  The explicit form of $h$ can be found using the fact that $g \in G
  \times \mathrm{U}(1)$ can be written as in
  Eq.~\eqref{eq:g-u(1)-decomposition},
  \begin{equation}
    g = k \, \exp \left[ \imath z \tanh \Theta  \left( T_C + T_{r+1}  \right) \right] \, ,  
  \end{equation}
  where $k \in G$ does not depend on $z$. Then left and right currents
  read:
  \begin{align}
    j &= \ee^{-\imath z \tanh \Theta \left( T_C + T_{r+1}  \right)} \left( k^{-1} \di k + \imath \tanh \Theta \di z \left( T_C + T_{r+1} \right) \right) \ee^{\imath z \tanh \Theta \left( T_C + T_{r+1}  \right)} \,,\\
    \overline j &= - \di k \, k^{-1} - \imath \tanh \Theta \di z
    \left( k T_C k^{-1} + T_{r+1} \right) \, .
  \end{align}
  The right-moving currents $\overline j$ do not depend on $z$;
  this is \emph{not} the case for $j$. If we choose
  \begin{equation}
    \label{eq:general-gauge-transform}
    h = \exp \left[ -\imath z \tanh \Theta \left( T_C + T_{r+1} \right)  \right] \, , 
  \end{equation}
  it is immediate to verify that
  \begin{equation}
    J' = -\Lambda(\zeta) \left( k^{-1} \di k + \imath \tanh \Theta \di z \left( T_C + T_{r+1} \right) \right)  - \imath \di z \tanh \Theta  \left( T_C + T_{r+1}  \right)
  \end{equation}
  only depends on the differential $\di z$ and can be T--dualized, resulting in
  \begin{equation}
    \widetilde J'= - \imath \left(\Lambda(\zeta) + 1 \right) \left( \st \di \wt z + \imath \tanh \Theta
    j_C \right) \tanh\Theta \left( T_C + T_{r+1} \right) -
  \Lambda(\zeta) k^{-1} \di k
  \end{equation}
\end{proof}

The presence of the zero mode $\wt J'^{(0)}$ is related to the fact
that $\wt J'$ cannot be understood in terms of a Noether current
resulting from an isometry of the T--dual metric. 
Note that in the T--dual model, we thus constructed an additional, non-local conserved  current $\wt J(\zeta)$ which does not stem from an isometry of the metric, but from a symmetry of the \textsc{eom}. The integrability of the T--dual model can thus be ascertained via the existence of an infinite family of conserved charges constructed from the conserved current which is T--dual to the Lax current of the original model.
$\wt {\overline J}(\zeta)$ on the other hand can be interpreted geometrically as a Noether current.

Observe that the flatness of $\wt J'(\zeta)$ implies the flatness of
$\wt J'^{(0)}$:
\begin{equation}
  \di \wt J'^{(0)} + \wt J'^{(0)} \wedge \wt J'^{(0)} = 0 \, . 
\end{equation}
This means that both $\wt J (\zeta)$ and $\wt J'^{(0)}$ can be
written in terms of Wilson lines $\wt W$ and $\wt W^{(0)}$:
\begin{align}
  \wt J' (\zeta)&= \wt W(\zeta)^{-1} \di \wt W(\zeta) \, , &   \wt J'^{(0)} &= (
  \wt W^{(0)})^{-1} \di \wt W^{(0)} \, ,
\intertext{where}
  \wt W(\zeta) &= \mathop{\mathrm{P}} \left\{ \exp \left[
      \int_{\mathcal{C}} \wt J'( \zeta) \right] \right\}, & \wt W^{(0)}
  &= \mathop{\mathrm{P}} \left\{ \exp \left[ \int_{\mathcal{C}} \wt
      J'^{(0)} \right] \right\} \, ,
\end{align}
and $\mathcal{C}$ is a path in $\Sigma$. Note that $\wt W^{(0)} $ is
the first term in the development in $\zeta$ of $\wt
W(\zeta)$:
\begin{equation}
  \wt W^{(0)} = \left. \wt W(\zeta) \right|_{\zeta = 0} \, . 
\end{equation}
The flatness of $\wt J'$ is all we need to define a one-parameter
family of conserved charges:
\begin{equation}
  \wt Q(t; \zeta) = \wt W(\infty, t | -\infty, t; \zeta) = \mathop{\mathrm{P}}\left\{ \exp \left[ \int_{-\infty}^\infty \wt J'_x(\xi, t; \zeta) \di \xi \right] \right\}\, ,
\end{equation}
which are in general \emph{non-local}. It is the existence of this infinite family of conserved charges which makes the model integrable.

These charges and the transformations they generate via Poisson brackets (which form a
$\widehat{\mathfrak{g}}$ algebra) can be understood in terms of
symmetries of the \textsc{eom} and a hierarchy of first order
equations (see Appendix~\ref{sec:deform-hier-squash}).

\section{Examples}
\label{sec:examples}

In this Section, we put the general formalism developed in
Section~\ref{sec:general-framework} into practice by explicitly
working out three examples. In Section~\ref{sec:SqS3}, we discuss the
example of the squashed three-sphere, a background appearing
frequently in the context of conformal field theory applications,
black holes in string theory, and the \AdS/\CFT correspondence (squashed
giant gravitons).  In an analogous manner, Section~\ref{sec:WAdS}
treats the warped anti--de Sitter space which has featured prominently
in the discussion of topologically massive gravity.  Differing
slightly from the other examples because of the T--duality being
performed in a non-compact direction,
Section~\ref{sec:example-thre-schr} spells out the Lax currents of the
Schroedinger spacetime $\Sch_3$, a background which has met with
sustained interest in the context of strongly coupled non-relativistic
quantum field theories.

\subsection{Example one: the squashed three-sphere}
\label{sec:SqS3}

As a first application of this general construction let us consider
the case of the squashed three-sphere.  In the simplest case, the
geometry of $\mathrm{SU}(2) \times \mathrm{U}(1) $ can be understood
as the fibration
\begin{equation}
  \begin{CD}
    S^1 \times S^1 @>>> S^3 \times S^1 \\
    @.      @VVV\\
    {} @. S^2
  \end{CD}
\end{equation}
where one of the directions in the torus fibration is the \emph{Hopf fiber}
in $S^3$. If we instead consider an $S^1$ sub-bundle $A$ of the torus, obtained as a rational
linear combination of the two $S^1$ directions above, we obtain the fibration
\begin{equation}
  \begin{CD}
    A @>>> S^3 \times S^1 \\
    @.      @VVV\\
    {} @. \SqS^3
  \end{CD}
\end{equation}
where $\SqS^3$ is the \emph{squashed three-sphere}.

It is convenient to choose a coordinate system in which the group
element $g \in \mathrm{SU}(2) \times \mathrm{U}(1)$ is written as
\begin{equation}
  g (\phi, \theta, \psi, y ) = \ee^{\imath \phi T_3 } \ee^{\imath \theta T_2} \ee^{\imath \psi T_3} \ee^{ 2 \imath y T_4} \, ,  
\end{equation}
where $\Theta $ is a real parameter and 
\begin{align}
  \label{eq:su2u1-generators}
  T_1 &= \frac{1}{2}  \begin{pmatrix} 0 & 1  \\ 1 & 0 \end{pmatrix} , &
  T_2 &= \frac{1}{2} \begin{pmatrix} 0 & -\imath  \\ \imath & 0 \end{pmatrix} , &
  T_3 &= \frac{1}{2} \begin{pmatrix} 1 & 0 \\ 0 & -1 \end{pmatrix} ,&
  T_4 &= \frac{1}{2\sinh \Theta} \begin{pmatrix} 1 & 0 \\ 0 & 1 \end{pmatrix}
\end{align}
are the generators of the algebra $\mathfrak{su}(2) \oplus
\mathfrak{u}(1)$:
\begin{align}
  \comm{T_1}{T_2} &= \imath T_3\,, &    \comm{T_2}{T_3} &= \imath T_1\,, &  
  \comm{T_3}{T_1} &= \imath T_2 \,,&  \comm{T_4}{T_a} &= 0\,,
\end{align}
with scalar product
\begin{equation}
  2 \Tr [ T_a T_b ] =
  \begin{cases}
    1 & \text{if $a=b=1,2,3$} \\
    \sinh^{-2} \Theta & \text{if $a=b=4$} \\
    0 & \text{if $a \neq b$}\,.
  \end{cases}
\end{equation}
Following~\cite{Orlando:2010ay} we introduce the variables $\alpha $
and $z$ as in Eq.~(\ref{eq:change-of-variables}):
\begin{align}
  \alpha &= \psi - 2 y\,, & z &= \frac{2 y}{\tanh \Theta} \, . 
\end{align}
The resulting metric on $\mathrm{SU}(2) \times \mathrm{U}(1)$ ($\di
s^2 = 2 \Tr [ \di g \di g^{-1} ]$) takes the form:
\begin{multline}
\label{eq:SU2U1-metric}
  \di s^2 =  \left[ \di \theta^2 + \sin^2 \theta \di \phi^2 +
    \frac{1}{\cosh^2 \Theta} \left(\di \alpha + \cos \theta \di \phi
    \right)^2 \right]  \\ + \left( \di z + \tanh \Theta
    \left( \di \alpha + \cos \theta \di \phi \right) \right)^2   \, ,
\end{multline}
which describes a fibration with fiber $z$, as in
Eq.~(\ref{eq:fibration-metric}) where $u^i = \set{\theta, \phi,
  \alpha}$ and $G_{ij}$ is the metric of a squashed
three-sphere. Performing the T--duality on $z$ leads to the condition
\begin{equation}
  \label{eq:T-duality-condition-SU2}
  \st \di \wt z = \di z + \tanh \Theta \left( \di \alpha + \cos
    \theta \di \phi \right),
\end{equation}
and substituting this into the action we obtain the metric in Eq.~\eqref{eq:ds2-SqG}:
\begin{equation}
\label{eq:SqS3-metric}
  \widetilde{\di s}^2 =  \left[ \di \theta^2 + \sin^2 \theta \di \phi^2 +
    \frac{1}{\cosh^2 \Theta} \left(\di \alpha + \cos \theta \di \phi
    \right)^2 \right] + \di \widetilde z^2 \, ,    
\end{equation}
where $\wt z$ is the dual variable. This is precisely the metric on
$\SqS^3 \times\, S^1$. Observe that by construction, the initial
$\mathrm{SU}(2) \times \mathrm{SU}(2) \times \mathrm{U}(1)$ isometry
group has been broken to $\mathrm{SU}(2) \times \mathrm{U}(1)^2$. The
corresponding Killing vectors are:
\begin{subequations}
  \begin{align}
    k{}^3 &= \del_\alpha \, , \\
    k{}^4 &= \del_{\wt z} \, , \\
    \overline k{}^1 &= \sin \phi \del_\theta + \cos \phi \frac{\cos
      \theta}{\sin \theta} \del_\phi - \frac{\cos \phi}{\sin \theta}
    \del_\alpha \, , \\
    \overline k{}^2 &= \cos \phi \del_\theta - \sin \phi \frac{\cos
      \theta}{\sin \theta}\del_\phi + \frac{\sin \phi}{\sin
      \theta} \del_\alpha \, , \\
    \overline k{}^3 &= \del_\phi \, .%
  \end{align}
\end{subequations}
The Killing vectors do not depend on the deformation parameter
$\Theta$ and are the same as for the initial three-sphere.

Our main result is that the initial symmetry can be restored and
promoted to affine $\widehat{\mathfrak{su}}(2) \oplus
\widehat{\mathfrak{su}}(2) \oplus \widehat{\mathfrak{u}}(1)$ thanks to
the presence of non-local charges that cannot be found via a standard
Noether construction.

In order to realize this symmetry explicitly, we start with the conserved currents
of the initial $\mathrm{SU}(2)^2 \times \mathrm{U}(1)$ principal chiral
model. In terms of variables $\set{\theta, \phi, \alpha, z}$, the
conserved currents $j$ and $\overline j$ are:
\begin{subequations}
\label{eq:SU2U1-currents}
  \begin{align}
    j^1 &= \imath \left( \cos ( \alpha + \tanh \Theta \, z)
      \sin \theta \di \phi - \sin ( \alpha + \tanh \Theta \, z ) \di
      \theta \right) \, , \\
    j^2 &= \imath \left( \sin ( \alpha + \tanh \Theta \, z)
      \sin \theta \di \phi + \cos ( \alpha + \tanh \Theta \, z ) \di
      \theta \right)  \, , \\
    j^3 &= \imath \left( \di \alpha + \cos \theta \di \phi +
      \tanh \Theta \di z \right)  \, , \\
    j^4 &= \imath  \tanh \Theta \di z \, ,
    \intertext{and} 
    \overline j^1 &= \imath \left( \cos \phi \sin
      \theta \di \alpha - \sin \phi \di \theta + \tanh \Theta \cos
      \phi \sin \theta \di z \right)  \, , \\
    \overline j^2 &= - \imath \left( \sin \phi \sin \theta \di
      \alpha + \cos \phi \di \theta + \tanh \Theta \sin \phi \sin
      \theta \di z \right)  \, , \\
    \overline j^3 &= - \imath \left( \cos \theta \di \alpha + \di
      \phi + \tanh \Theta \cos \theta \di z \right)  \, , \\
    \overline j^4 &= - \imath \tanh \Theta \di z  \, ,
  \end{align}
\end{subequations}
where $j = g^{-1} \di g$ and $\overline j = - \di g g^{-1}$ have been
decomposed on the generators $T_a$ of the algebra introduced in
Eq.~(\ref{eq:su2u1-generators})\footnote{Since the currents are maps
  $j : \Sigma \to \mathfrak{g}$, they can be decomposed on a basis
  $\set{T_a}$ of $\mathfrak{g}$,
  \begin{align}
    j (x,t)&= \sum_{a=1}^{\dim G } j^a (x,t) T_a, & 
    \overline j (x,t) &= \sum_{a=1}^{\dim G } \overline j^a (x,t) T_a \, .
  \end{align}}.

We can now apply the procedure described in
Section~\ref{sec:integr-squash-groups} to obtain the T--dual currents
for the model with the metric in Eq.~(\ref{eq:SqS3-metric}). In order
to impose the condition in Eq.~(\ref{eq:T-duality-condition-SU2}) ($
\st \di \wt z = \di z + \tanh \Theta \left( \di \alpha + \cos
  \theta \di \phi \right) $) on the $\mathfrak{su}(2)
\oplus \mathfrak{su}(2) \oplus \mathfrak{u}(1)$ Lax currents, we first have to perform a gauge
transformation since $j^3$ (appearing in the T--duality
transformation) does not commute with the currents $j^1 $ and $j^2$ in
Eq.~(\ref{eq:SU2U1-currents}), as one can see in this coordinate system
since they depend explicitly on the variable $z$. Instead of
$J(\zeta)$, consider the flat current
\begin{equation}
  J'(\zeta) = h^{-1} J(\zeta) h + h^{-1} \di h \, ,
\end{equation}
where\footnote{Note that this transformation differs from the one in
  Eq.~\eqref{eq:general-gauge-transform} and it has been chosen
  because it leads to simpler expressions for the currents in the new
  gauge.}
\begin{equation}
  h = \exp \left[ -\imath \left( \alpha + \tanh \Theta \, z \right) T_3 \right] \, .
\end{equation}
Explicitly,
\begin{subequations}
  \begin{align}
   J'{}^1(\zeta) &= -\imath \Lambda(\zeta) \sin \theta \di
    \phi\,,
    \\
    J'{}^2(\zeta) &= - \imath \Lambda(\zeta) \di \theta \,,\\
    {J'}^3(\zeta) &= - \imath \left[ \left( \di \alpha +
        \tanh \Theta \di z \right) + \Lambda(\zeta) \left( \di
        \alpha + \cos \theta \di \phi + \tanh \Theta \di z
      \right) \right]\,, \\
    J'{}^4(\zeta) &= - \imath \tanh \Theta \Lambda(\zeta)  \di z\,.
  \end{align}
\end{subequations}
We can now impose the T--duality condition in
Eq.~(\ref{eq:T-duality-condition-SU2}),
\begin{equation}
  \st \di \wt z = \di z + \tanh \Theta \left( \di \alpha + \cos \theta \di \phi \right)\,,
\end{equation}
and find
\begin{subequations}
  \begin{align}
    &\wt J'{}^1(\zeta) = - \imath \Lambda(\zeta) \sin
    \theta \di \phi\,, \\
    &\wt J'{}^2(\zeta) =  - \imath \Lambda(\zeta) \di \theta\,, \\
    &\wt J'{}^3(\zeta) = - \imath \left[ \left( 1 + \Lambda(\zeta) \right) \left( \frac{\di \alpha + \cos \theta \di \phi}{\cosh^2 \Theta} + \tanh \Theta \st \di \wt z \right) - \cos \theta \di \phi \right]\,,\\
  &\wt J'{}^4(\zeta) = - \imath  \tanh \Theta
     \Lambda(\zeta) \left( \st \di \wt z - \tanh
      \Theta \left( \di \alpha + \cos \theta
        \di \phi \right) \right)\,, \\
    \intertext{and}
    \begin{split}
      \wt{ \overline J}{}^1 (\zeta) = - \imath \Lambda(\zeta) \Big[
      \frac{1}{\cosh^2 \Theta} \cos \phi \sin \theta \di \alpha -
       \sin \phi \di\theta+ \tanh^2 \Theta \cos \phi \sin \theta \cos
        \theta \di \phi +{} \\ {}+ \tanh \Theta \cos \phi \sin
      \theta \st \di \wt z \Big]\,,
    \end{split} \\
    \begin{split}
      \wt{ \overline J}{}^2 (\zeta) = \imath \Lambda(\zeta) \Big[
      \frac{1}{\cosh^2 \Theta} \sin \phi \sin \theta \di \alpha +
       \cos \phi \di \theta - \tanh^2 \Theta \sin \phi \sin \theta \cos
        \theta \di \phi + {} \\ {} + \tanh \Theta \sin \phi \sin
      \theta \st \di \wt z \Big]\,,
    \end{split} \\
    &\wt{ \overline J}{}^3(\zeta) = \imath \Lambda(\zeta) \left[
      \frac{1}{\cosh^2 \Theta} \cos \theta \di \alpha + \left( 1 -
        \tanh^2 \Theta \cos^2 \theta \right) \di \phi + \tanh
      \Theta
      \cos \theta \st \di \wt z \right] \,,\\
    &\wt{ \overline J}{}^4 (\zeta)= \imath  \tanh \Theta
    \Lambda(\zeta) \left( \st \di \wt z - \tanh \Theta \left( \di
        \alpha + \cos \theta \di \phi \right) \right)\,.
  \end{align}
\end{subequations}
Note that the power series expansion in $\zeta$ of the current $\wt
J'(\zeta)$ has a zero-order component:
\begin{equation}
  \wt J'{}^{(0)} = h^{-1} \di h \, = - \imath \left( \di \alpha +
        \tanh \Theta \st \di \wt z-\tanh^2 \Theta \left( \di \alpha + \cos \theta \di \phi  \right) \right)T_3 \, . 
\end{equation}
Note also that while $\wt J'(\zeta)$ generates non-local charges which
stems from a symmetry of the \textsc{eom} which is \emph{not} an
isometry of the metric, $\wt{\overline J}'(\zeta)$ descends from an
isometry of the metric and has thus a geometrical interpretation as a
Noether current. 

The $S^2 \times S^1$ geometry and currents can be found in the limit
$\Theta \to \infty$.

\subsection{Example two: warped anti--de Sitter space}
\label{sec:WAdS}

In a completely analogous fashion to the last example, we can instead
start from the group $\mathrm{SL}_2( \setR) \times \mathrm{U}(1)$.
Also in this case we will see how the $\mathrm{SL}_2(\setR)^2 \times
\mathrm{U}(1)$ isometry is promoted to affine
$\widehat{\mathfrak{sl}_2}(\setR) \oplus
\widehat{\mathfrak{sl}_2}(\setR) \oplus \widehat{\mathfrak{u}}(1)$.

It is convenient to choose a coordinate system in which the group
element $g \in \mathrm{SL}_2(\setR) \times \mathrm{U}(1)$ is written as
\begin{equation}
  g (\tau, \omega, \sigma, y ) = \ee^{- \tau T_1 } \ee^{\omega T_3} \ee^{\imath \sigma T_2} \ee^{ 2 \imath y T_4} \, ,  
\end{equation}
where $\Theta $ is a real parameter and 
\begin{align}
  \label{eq:sl2u1-generators}
  T_1 &= \frac{1}{2} \begin{pmatrix} 0 & \imath \\ \imath & 0 \end{pmatrix}  , &
  T_2 &= \frac{1}{2} \begin{pmatrix} \imath & 0 \\ 0 & - \imath \end{pmatrix}  , &
  T_3 &= \frac{1}{2} \begin{pmatrix} 0 & \imath\\ -\imath & 0 \end{pmatrix}  ,&
  T_4 &= \frac{1}{2 \sinh \Theta} \begin{pmatrix} 1 & 0 \\ 0 & 1 \end{pmatrix}
\end{align}
are the generators of the algebra $\mathfrak{gl}_2$:
\begin{align}
  \comm{T_1}{T_2} &= - \imath T_3\,, &    \comm{T_2}{T_3} &= \imath T_1\,, &  
  \comm{T_3}{T_1} &= \imath T_2\,, &  \comm{T_4}{T_a} &= 0\,,
\end{align}
 with scalar product
\begin{equation}
  - 2 \Tr [ T_a T_b ] =
  \begin{cases}
    1 & \text{if $a = b = 1,2$} \\
    -1 & \text{if $a = b = 3$} \\
    \sinh^{-2} \Theta & \text{if $a = b = 4$} \\
    0 & \text{if $a \neq b$.}
  \end{cases}
\end{equation}
Following~\cite{Orlando:2010ay} we introduce the variables $\beta $ and $z$:
\begin{align}
  \beta &= \sigma - 2 y \,, & z &= \frac{2 y}{\tanh \Theta} \, . 
\end{align}
The resulting metric on $\mathrm{SL}_2(\setR) \times \mathrm{U}(1)$ takes
the form
\begin{multline}
\label{eq:SL2U1-metric}
  \di s^2 =  \left[ \di \omega^2 - \cosh^2 \omega \di \tau^2 +
    \frac{1}{\cosh^2 \Theta} \left(\di \beta + \sinh \omega \di \tau
    \right)^2 \right] \\ + \left( \di z + \tanh \Theta
    \left( \di \beta + \sinh \omega \di \tau \right) \right)^2   \, ,
\end{multline}
which describes a fibration with fiber $z$, as in
Eq.~(\ref{eq:fibration-metric}) where $u^i = \set{\omega, \tau,
  \beta}$, and $G_{ij}$ is the metric of a \emph{warped anti--de Sitter
space}. Performing the T--duality on $z$ leads to the condition
\begin{equation}
  \label{eq:T-duality-condition-SL2}
  \st \di \wt z = \di z + \tanh \Theta \left( \di \beta + \sinh
    \omega \di \tau \right).
\end{equation}
Substituting this into the action we obtain the following metric:
\begin{equation}
  \label{eq:WAdS3-metric}
  \wt {\di s}^2 = \left[ \di \omega^2 - \cosh^2 \omega \di \tau^2 +
    \frac{1}{\cosh^2 \Theta} \left(\di \beta + \sinh \omega \di \tau
    \right)^2 \right] + \di \wt z^2 \, ,    
\end{equation}
where $\wt z$ is the dual variable. This is precisely the metric on
$\WAdS_3 \times S^1$, where by construction the initial
$\mathrm{SL}_2(\setR) \times \mathrm{SL}_2(\setR) \times \mathrm{U}(1)$ isometry group has
been broken to $\mathrm{SL}_2(\setR) \times \mathrm{U}(1)^2$.
The corresponding Killing vectors (that do not depend on the
deformation parameter $\Theta$) are:
\begin{subequations}
  \begin{align}
    k{}^2 &= \del_\beta \, , \\
    k{}^4 &= \del_{\wt z} \, ,\\
    \overline k{}^1 &= \del_\tau \, , \\
    \overline k{}^2 &= \cos \tau \tanh \omega \del_\tau + \frac{\cos
      \tau}{\cosh \omega} \del_\beta + \sin \tau \del_\omega \, , \\
    \overline k{}^3 &= \sin \tau \tanh \omega \del_\tau + \frac{\sin
      \tau}{\cosh \omega} \del_\beta - \cos \tau \del_\omega \, . %
  \end{align}
\end{subequations}

Once more, the initial symmetry can be restored and promoted to affine
$\widehat{\mathfrak{sl}}_2 \oplus \widehat{\mathfrak{sl}}_2 \oplus
\widehat{\mathfrak{u}}(1)$ thanks to the presence of a non-local
current that cannot be found via a standard Noether construction.

In the following, we will construct this symmetry explicitly. In terms
of the variables $\set{\omega, \tau, \beta, z}$, the conserved
currents before the T--duality $j$ and $\overline j$ are given by
\begin{subequations}
\label{eq:SL2U1-currents}
  \begin{align}
    j^1 &= - \cosh( \beta + \tanh \Theta \, z ) \cosh \omega \di \tau
    + \sinh( \beta + \tanh \Theta \, z) \di \omega\,, \\
    j^2 &= \imath \left( \di \beta + \sinh \omega \di \tau + \tanh
      \Theta \di z \right)\,, \\
    j^3 &= \cosh( \beta + \tanh \Theta \, z ) \di \omega - \sinh (
    \beta + \tanh \Theta \, z ) \cosh  \omega \di \tau\,, \\ 
    j^4 &= \imath \tanh \Theta \di z \, ,
    \intertext{and} 
    \overline j^1 &= \di \tau - \sinh \omega \di \beta - \tanh \Theta
    \sinh \omega \di z\,, \\
    \overline j^2 &= - \imath \left( \cos \tau \cosh \omega \di \beta +
      \sin \tau \di \omega + \cos \tau \cosh \omega \tanh \Theta \di
    z \right)\,, \\
    \overline j^3 &= -\cos \tau \di \omega + \sin \tau \cosh \omega \di
    \beta + \sin \tau \cosh \omega \tanh \Theta \di z\,, \\
    \overline j^4 &= - \imath \tanh \Theta \di z  \, ,
  \end{align}
\end{subequations}
where $j = g^{-1} \di g$ and $\overline j = - \di g g^{-1}$ have been
decomposed on the generators $T_a$ of the algebra introduced in
Eq.~(\ref{eq:sl2u1-generators}).

We can again apply the procedure described above to the T--duality
transformation that leads to the $\WAdS$ metric. In order to impose
the condition in Eq.~(\ref{eq:T-duality-condition-SL2}) ($ \st \di \wt
z = \di z + \tanh \Theta \left( \di \beta + \sinh \omega \di
  \tau \right) $) to the Lax currents for $\mathfrak{sl}_2 \oplus
\mathfrak{sl}_2 \oplus \mathfrak{u}(1)$, we first have to perform a gauge transformation
\begin{equation}
  J'(\zeta) = h^{-1} J(\zeta) h + h^{-1} \di h \, ,
\end{equation}
where
\begin{equation}
  h = \exp \left[ -\imath \left( \beta + 
      \tanh \Theta \, z  \right)  T_2 \right] \, .
\end{equation}
Explicitly,
\begin{subequations}
  \begin{align}
    J'{}^1(\zeta) &= \Lambda(\zeta) \cosh \omega \di \tau \,,\\
    J'{}^2(\zeta) &= - \imath \left[ \left( \di \beta + \tanh \Theta \di
        z \right) + \Lambda(\zeta) \left( \di \beta + \sinh \omega
        \di \tau + \tanh \Theta \di z \right) \right]\,,\\
    {J'}^3(\zeta) &= - \Lambda(\zeta) \di \omega\,, \\
    J'{}^4(\zeta) &= - \imath \tanh \Theta \Lambda(\zeta) \di z.
  \end{align}
\end{subequations}
We can now impose the T--duality condition in
Eq.~(\ref{eq:T-duality-condition-SL2})
\begin{equation}
  \st \di \wt z = \di z + \tanh \Theta \left( \di \beta + \sinh \omega \di \tau \right)
\end{equation}
and find
\begin{subequations}
  \begin{align}
    &\wt J'{}^1(\zeta) = \Lambda(\zeta) \cosh \omega \di \tau\,, \\
    &\wt J'{}^2(\zeta) = - \imath \left[ \left( 1 + \Lambda(\zeta) \right) \left( \frac{\di \beta + \sinh \omega \di \tau}{\cosh^2 \Theta} + \tanh \Theta \st \di \wt z \right) - \sinh \omega \di \tau \right]\,,\\
    &\wt J'{}^3(\zeta) = - \Lambda(\zeta) \di \omega\,, \\
    &\wt J'{}^4(\zeta) = - \imath \tanh \Theta \Lambda(\zeta) \left( \st \di \wt z - \tanh \Theta
      \left( \di \beta + \sinh \omega \di \tau \right)\right)\,,
    \intertext{and}
    &\wt{ \overline J}{}^1 (\zeta) = \Lambda(\zeta) \left[ -\left( 1 + \tanh^2 \Theta \sinh^2
        \omega \right) \di \tau + \frac{1}{\cosh^2 \Theta} \sinh
      \omega \di \beta + \tanh \Theta \sinh \omega \st \di \wt z
    \right]\,, \\
    \begin{split}
      \wt{ \overline J}{}^2 (\zeta) = \imath \Lambda(\zeta) \Big[
        \frac{1}{\cosh^2 \Theta} \cos \tau \cosh \omega \di \beta +
        \sin \tau \di \omega - \cos \tau \cosh \omega \sinh \omega
        \tanh^2 \Theta \di \tau + {} \\ {} + \cos \tau \cosh \omega \tanh
        \Theta \st \di \wt z \Big] \,,
    \end{split} \\
    \begin{split}
      \wt{ \overline J}{}^3 (\zeta) =- \Lambda(\zeta) \Big[
        \frac{1}{\cosh^2 \Theta} \sin \tau \cosh \omega \di \beta -
        \cos \tau \di \omega - \sin \tau \cosh \omega \sinh \omega
        \tanh^2 \Theta \di \tau + \\ + \sin \tau \cosh \omega \tanh
        \Theta \st \di \wt z \Big] \,,
    \end{split} \\
    &\wt{ \overline J}{}^4 (\zeta) =  \imath \tanh \Theta \Lambda(\zeta)
    \left( \st \di \wt z - \tanh \Theta \left( \di \beta + \sinh
        \omega \di \tau \right) \right)\,.
  \end{align}
\end{subequations}
Note that also the power series expansion in $\zeta$ of the current
$\wt J'(\zeta)$ has a zero-order component:
\begin{equation}
  \wt J'{}^{(0)} = h^{-1} \di h =  \imath\left( \di \beta + 
      \tanh \Theta  \st \di \wt z-\tanh^2 \Theta \left( \di \beta + \sinh \omega \di \tau \right)  \right)  T_2 \, . 
\end{equation}
Just like in the previous example, the current $\wt{\overline
  J}(\zeta)$ can be regarded as a Noether current, while this is not
the case for $\wt J'(\zeta)$.

The limit $\Theta \to \infty$ describes the currents in the geometry
$\AdS_2 \times S^1$.

\subsection{Example three: Schroedinger spacetime}
\label{sec:example-thre-schr}

Cartan subgroups of non-compact groups are in general not related by
inner automorphisms. This leaves
more freedom in the choice of the direction in which to perform the
T--duality. The \emph{Schroedinger spacetime} is another
example of geometry that can be obtained starting from $\AdS_3$. The
main difference with respect to the previous example is that the
T--duality is performed in a \emph{non-compact} direction. The general form of a $(d+1)$-dimensional
Schroedinger spacetime is given by
\begin{equation}
  \label{Schgen} 
  ds^2=-\frac{b^2 \di{x^{-}}^2}{r^4}+\frac{2\di x^{-}\di x^{+} + \di x^i \di x^i + \di r^2}{r^2}
\end{equation}
with $i=1,\cdots d-2$ . A list of all the isometries of this metric
can be found in~\cite{Guica:2010sw}. It was initially believed that
this background could be holographically dual to a critical
non-relativistic system in $(d-1)$ spacetime dimensions having the
same symmetries. It was later shown that one requires the coordinate
$x^{+}$ to be a compact null direction, which introduces a series of
complications when looking at quantum corrections
\cite{Maldacena:2008wh}. Nonetheless, an holographic dictionary has
been established for the case in which $x^{+}$ is non-compact and it
has been shown that the space described by Eq.~(\ref{Schgen}) is dual
to a $d$-dimensional QFT which is non-local in the $x^{+}$ direction.

We will look at $d=2$ which corresponds to $\Sch_3$. We start by
considering a choice of coordinates where the group element $g \in
\mathrm{GL}_2(\setR)$ is written as
\begin{equation}
  g ( r, u^+, x^-, z ) = \ee^{ \imath x^- \left( T_3 - T_1 \right)}
  \ee^{2 \imath \log r \, T_2} \ee^{-\imath u^+ \left( T_3 + T_1 \right)} \ee^{2 \imath z T_4},
\end{equation}
where
\begin{align}
  \label{eq:sl2u1-generators}
  T_1 &= \frac{1}{2} \begin{pmatrix} 0 & \imath \\ \imath & 0 \end{pmatrix}  , &
  T_2 &= \frac{1}{2} \begin{pmatrix} \imath & 0 \\ 0 & - \imath \end{pmatrix}  , &
  T_3 &= \frac{1}{2} \begin{pmatrix} 0 & \imath\\ -\imath & 0 \end{pmatrix}  ,&
  T_4 &= \frac{1}{2} \begin{pmatrix} 1 & 0 \\ 0 & 1 \end{pmatrix}
\end{align}
are the generators of the algebra $\mathfrak{gl}_2$, with scalar
product
\begin{equation}
  - 2 \Tr [ T_a T_b ] =
  \begin{cases}
    1 & \text{if $a = b = 1,2,4$} \\
    -1 & \text{if $a = b = 3$} \\
    0 & \text{if $a \neq b$.}
  \end{cases}
\end{equation}
In this case we do not introduce the parameter $\Theta$, since we are
interested in a non-compact direction and this parameter could be
reabsorbed by a coordinate redefinition.  Let us now introduce the coordinate
$x^+$:
\begin{equation}
  u^+ = x^+ + z \, .
\end{equation}
The resulting metric on $\mathrm{GL}_2(\setR) $ takes
the form
\begin{equation}
  \label{eq:GL2-metric-Schro}
  \di s^2 =  \left[ \frac{\di r^2}{r^2} + \frac{\di x^+ \di x^-}{r^2} - \frac{\left( \di x^- \right)^2 }{4 r^4} \right] +  \left( \di z +\frac{\di x^-}{2 r^2} \right)^2  ,
\end{equation}
which describes a fibration with fiber $z$, as in
Eq.~(\ref{eq:fibration-metric}) where $u^i = \set{r, x^+, x^-}$, and
$G_{ij}$ is the metric of a three-dimensional Schroedinger
  spacetime. Performing the T--duality on $z$ leads to the
condition
\begin{equation}
  \label{eq:T-duality-condition-Schro}
  \st \di \wt z =  \di z +\frac{\di x^-}{2 r^2} \, .
\end{equation}
Substituting this into the action we obtain the metric
\begin{equation}
  \label{eq:Schro-metric}
  \wt {\di s}^2 = \left[ \frac{\di r^2}{r^2} + \frac{\di x^+ \di x^-}{r^2} - \frac{\left( \di x^- \right)^2 }{4 r^4} \right] + \di \wt z^2,
\end{equation}
where $\wt z$ is the dual variable. This is precisely the metric on
$\Sch_3 \times S^1$, where by construction the initial
$\mathrm{SL}_2(\setR) \times \mathrm{SL}_2(\setR) \times \setR$
isometry group has been broken to $\mathrm{SL}_2(\setR) \times
\setR^2$.  The corresponding Killing vectors are:
\begin{subequations}
  \begin{align}
    k{}^- &= \del_{x^+} \, , &
    k{}^4 &= \del_{\wt z} \, , \\
    \overline k{}^- &= r x^- \del_r - r^2 \del_{x^+} + \left( x^-
    \right)^2 \del_{x^-} \, , & 
    \overline k{}^2 &= r \del_r + 2 x^- \del_{x^-} \, , &
    \overline k{}^+ &= \del_{x^-} \, . %
  \end{align}
\end{subequations}

Once more, the initial symmetry can be restored and promoted to affine
$\widehat{\mathfrak{sl}}_2 \oplus \widehat{\mathfrak{sl}}_2 \oplus
\widehat{\mathfrak{u}}(1)$ thanks to
the presence of a non-local current that cannot be found via a
standard Noether construction.

Before we move on to the construction of the currents it is useful to
comment on the norms of the explicit Killing vectors for the metrics
obtained via the change of variables and T--duality.
In the initial metric $\di s^2 \propto \Tr[ \di g \di g^{-1}]$, we have
\begin{align}
  \norm{\del_{u^+}}^2 &= 0 \, , & \norm{\del_{x^-}}^2 &= 0 \,, &
  \norm{\del_z}^2 &= 1 \, .
\end{align}
After the change of coordinates $u^+ = x^+ + z$, the variable $x^+$
describes a null direction and $z$ remains space-like:
\begin{align}
  \norm{\del_{x^+}}^2 &= 0 \, , & \norm{\del_{x^-}}^2 &= 0 \,,&
  \norm{\del_z}^2 &= 1 \, .
\end{align}
The T--duality changes the nature of $x^-$, which in the metric in
Eq.~(\ref{eq:GL2-metric-Schro}) is time-like:
\begin{align}
  \norm{\del_{x^+}}^2 &= 0 \, , & \norm{\del_{x^-}}^2 &= -
  \frac{1}{4 r^4} \,, &
  \norm{\del_{\wt z}}^2 &= 1 \, .
\end{align}
This can be changed to $\norm{\del_{x^-}}^2 = 1/(4 r^4)$ via a
double analytic continuation $\left( x^+, x^- \right) \mapsto \left( \imath
x^+, - \imath x^- \right)$ which does not affect the $(-,+,+,+)$
signature of the metric.

Let us now move to the currents. In terms of the variables $\set{r,
  x^+, x^-, z}$, the conserved currents before the T--duality $j$ and
$\overline j$ are given by
\begin{subequations}
\label{eq:Sch-currents}
  \begin{align}
    j^- &= - \frac{2 \imath \di x^-}{r^2} \,, \\
    j^2 &= 2 \imath \left( \frac{\di r}{r} + \frac{z \di x^-}{r^2} + \frac{x^+ \di x^-}{r^2} \right) ,\\
    j^+ &= 2 \imath \left( \frac{2 z + 2 x^+}{r} \di r + \frac{\left( z + x^+ \right)^2}{r^2} \di x^- - \di z - \di x^+ \right) , \\
    j^4 &= 2 \imath \di z\,, \\
    \intertext{and} 
    \overline j{}^- &= -2 \imath \left( \frac{2 x^-}{r} \di r - \di x^- + \frac{\left( x^- \right)^2}{r^2} \left( \di z + \di x^+ \right)  \right) , \\
    \overline j{}^2 &= -2 \imath \left( \frac{\di r}{r} + \frac{x^-}{r^2} \di z + \frac{x^-}{r^2} \di x^+ \right), \\
    \overline j{}^+ &= 2 \imath \left(\frac{\di z}{r^2} + \frac{\di x^+}{r^2} \right) ,\\
    \overline j{}^4 &= - 2 \imath \di z\,,
  \end{align}
\end{subequations}
where $j^{\pm} = j^1 \pm j^3$.

Using a by now familiar procedure, in order to impose the condition in
Eq.~(\ref{eq:T-duality-condition-Schro}) ($ \st \di \wt z = \di z +\di
x^-/(2 r^2) $) on the Lax currents for $\mathfrak{sl}_2 \oplus
\mathfrak{sl}_2 \oplus \mathfrak{u} (1)$, we first have to perform a
gauge transformation
\begin{equation}
  J'(\zeta) = h^{-1} J(\zeta) h + h^{-1} \di h \, ,
\end{equation}
where
\begin{equation}
  h = \exp \left[\imath \left( x^+ + z \right)  \left( T_1 + T_3 \right) \right] \, .
\end{equation}
Explicitly,
\begin{subequations}
  \begin{align}
    J'{}^-(\zeta) &= 2 \imath \Lambda(\zeta) \frac{\di x^-}{r^2} \, , \\
    J'{}^2(\zeta) &= - 2 \imath \Lambda( \zeta ) \frac{\di r}{r} \, , \\
    J'{}^+(\zeta) &= 2 \imath \left( 1 + \Lambda(\zeta) \right) \left( \di z + \di x^+  \right) \, ,\\
    J'{}^4(\zeta) &= - 2 \imath \Lambda(\zeta) \di z \, .
  \end{align}
\end{subequations}
We can now impose the T--duality condition in
Eq.~(\ref{eq:T-duality-condition-Schro}),
\begin{equation}
  \st \di \wt z = \di z + \frac{\di x^-}{2 r^2} \,,
\end{equation}
and find
\begin{subequations}
  \begin{align}
    \wt J'{}^-(\zeta) &= 2 \imath \Lambda(\zeta) \frac{\di x^-}{r^2} \, , \\
    \wt J'{}^2(\zeta) &= - 2 \imath \Lambda( \zeta ) \frac{\di r}{r} \, , \\
    \wt J'{}^+(\zeta) &= 2 \imath \left( 1 + \Lambda(\zeta) \right) \left( \st \di \wt z - \frac{\di x^-}{2 r^2} + \di x^+  \right) \, ,\\
    \wt J'{}^4(\zeta) &= - 2 \imath \Lambda(\zeta) \left( \st \di \wt z - \frac{\di x^-}{2 r^2}  \right), \\
    \intertext{and}
    \wt{\overline J}{}^-(\zeta) &= 2 \imath \Lambda(\zeta) \left( \frac{2 x^-}{r} \di r - \di x^- + \frac{\left( x^- \right)^2}{r^2} \left(  \st \di \wt z - \frac{\di x^-}{2 r^2} + \di x^+ \right)  \right) \, , \\
    \wt{\overline J}{}^2(\zeta) &= 2 \imath \Lambda(\zeta) \left( \frac{\di r}{r} + \frac{x^-}{r^2} \di z + \frac{x^-}{r^2} \di x^+ \right) \, , \\
    \wt{\overline J}{}^+(\zeta) &= -\frac{2\imath}{r^2} \Lambda(\zeta) \left( \st \di \wt z - \frac{\di x^-}{2 r^2} + \di x^+ \right) \, , \\
    \wt{\overline J}{}^4(\zeta) &= 2 \imath \Lambda(\zeta)\left( \st \di \wt z - \frac{\di x^-}{2 r^2} \right) \, .
  \end{align}
\end{subequations}
Just like in the previous example, the current $\wt{\overline
  J}(\zeta)$ can be regarded as a Noether current, while this is not
the case for $\wt J'(\zeta)$.

\section{Conclusions}
\label{sec:conclusions}

From a world-sheet point of view, T--duality is a linear
transformation of the components of the currents. It preserves the
integrable structure of the original model, at least at the classical
level, and can thus be used to generate new integrable sigma models. 

This mechanism is so powerful that it has allowed us to treat in a unified
way three different examples, which emerge naturally as target space backgrounds 
in string theory and that can be used to further study certain types of black holes
and through holography, non-relativistic quantum field theories.

\bigskip

Inspired by the fact that backgrounds containing squashed spheres can
be obtained via T--duality from an $\AdS_3 \times S^3$ background and
that the corresponding sigma model is integrable, we showed by direct
computation of the Lax pairs and the infinite set of conserved charges
how the integrability emerges in the T--dual model for the cases of
the squashed three-sphere, warped $\AdS$ and the Schroedinger
spacetime.

It is very interesting that despite the fact that the isometry group
of the T--dual model is only a subgroup of the original isometry
group, the T--dual currents lead to the full
$\mathfrak{g}\oplus\mathfrak{g}$ symmetry, realizing the ``hidden''
symmetry just as in the cases discussed in~\cite{Ricci:2007eq}. In the
T--dual model, the symmetry group can be promoted to an affine
symmetry due to the emergence of non-local charges that cannot be
obtained via a Noether construction. In this way we also generalize
the hidden Yangian algebra found in~\cite{Kawaguchi:2010jg}.

It should be remarked that our analysis is purely classical, which is
consistent with the fact that the algebra of the symmetries is a loop
algebra (\emph{i.e.} there is no central term). This changes once
quantum corrections are taken into account. Nevertheless we expect the
integrable structure to be preserved in the full supersymmetric sigma
model.

\bigskip

The natural next step is to extend the analysis performed here to the
case in which both RR fluxes and fermions are turned on. It would be
important to understand whether the original symmetries can be
realized in a non-local way in an analogous fashion and whether the
promotion to the affine symmetry can still be realized. This will be
the subject of a forthcoming publication~\cite{Reffert:2010fu}.

\subsection*{Acknowledgements}

We would like to thank Arkady Tseytlin for inspiring discussions and
detailed comments on the manuscript. We furthermore would like to
thank Konstadinos Sfetsos for correspondence. Moreover, D.O. and
S.R. would like to thank the participants of the IPMU string theory
group meetings for stimulating discussions. D.O. would like to thank
Io Kawaguchi and Kentaroh Yoshida for collaboration on a related
topic.  The research of D.O. and S.R. was supported by the World
Premier International Research Center Initiative (WPI Initiative),
MEXT, Japan. L.I.U. acknowledges the support of an STFC Postdoctoral
Fellowship.

\newpage
\appendix

\section{Hierarchy of variations for the principal chiral model}
\label{sec:hier-vari-princ}

The principal chiral model has an explicit $G\times
G$ symmetry. The infinitesimal version of this symmetry implies that
its action is invariant under
\begin{equation}
  \delta g = \epsilon g \quad \text{and} \quad \delta g=g\overline\epsilon,\quad \di\epsilon=0,\ \di\overline\epsilon=0\,,
\end{equation}
where $\epsilon $ and $\overline \epsilon$ are constants $\epsilon,
\overline \epsilon \in \mathfrak{g}$. Following~\cite{Schwarz:1995td}
we can rewrite the left action $G_L$, $\delta g=\epsilon g$ in terms
of a function $\eta: \Sigma \to \mathfrak{g}$:
\begin{equation}
  \delta g(x,t) = \epsilon g(x,t) = g(x,t) \eta(x,t) = \delta_\eta g(x,t) \, , \quad \eta(x,t) = g^{-1}(x,t)\epsilon g(x,t) \, .
\end{equation}
The function $\eta$ is \emph{not} constant on $\Sigma$ and this symmetry
should not be confused with $G_R$.

The variation of the current $j=g^{-1}\di g$ under $\delta_\eta$ is
\begin{equation}
  \delta_\eta j= \di \eta + [j,\eta]=\nabla_j \eta \, .
\end{equation}
We thus conclude that $\delta_\eta$ is a symmetry of the action if
$\eta$ is covariantly constant with respect to $j$:
\begin{equation}
  \nabla_j \eta (x,t)  = 0 \, .  
\end{equation}

In the spirit of integrability, instead of $\eta$, we  introduce a
one-parameter family of Lie algebra-valued functions $\eta_\zeta :
\Sigma \to \mathfrak{g}$ which are covariantly constant with respect
to the Lax current $J(\zeta)$,
\begin{equation}
  \nabla_J \eta_\zeta= \di\eta_\zeta+[J(\zeta), \eta_\zeta]=0.
\end{equation}
It follows that
\begin{multline}
  \nabla_j \eta_\zeta = \nabla_J \eta_\zeta + \comm{j - J(\zeta)}{\eta_\zeta}= \comm{\left( 1 + \Lambda(\zeta) \right) j }{\eta_\zeta}= \frac{1}{\zeta}\comm{\st \Lambda(\zeta) j}{\eta_\zeta} = - \frac{1}{\zeta}\comm{\st J(\zeta)}{\eta_\zeta}  \\ = \frac{1}{\zeta} \st \di\eta_\zeta \, ,
\end{multline}
where we used the property in Eq.~\eqref{eq:star-Lambda}.
This implies that the \textsc{eom} $\di\st j=0$ are conserved
under the variation $\delta \eta_\zeta$:
\begin{equation}
  \delta_{\eta_\zeta} ( \di \ast j ) = \di \ast (\delta_{\eta_\zeta} j )= \di \ast ( \nabla_j \eta_\zeta) =  \frac{1}{\zeta} \di \ast ( \ast \di \eta_{\zeta} ) = 0 \, .
\end{equation}
We thus have found a one-parameter family of symmetries of the
\textsc{eom}. The condition $\nabla_J{\eta_\zeta}=0$ can be
rewritten in terms of a hierarchy of first order equations. Expand
$\eta_\zeta$ in powers of $\zeta$:
\begin{equation}
  \eta_\zeta = \sum_{n=0}^\infty\zeta^n\eta^{(n)} \, .
\end{equation}
Then $\eta_\zeta $ is covariantly constant with respect to $J(\zeta) =
- \Lambda(\zeta) j $ if and only if
\begin{equation}
  \di \eta_\zeta = \comm{\Lambda(\zeta)j}{\eta_\zeta} 
\end{equation}
and, order by order
\begin{subequations}
  \label{eq:PCM-hierarchy}
  \begin{align}
    \di\eta^{(0)} &= 0,\\*
    \di\eta^{(1)} &= [\st j,\eta^{(0)}],\\*
    \di\eta^{(2)} &= [\st j,\eta^{(0)}]+[j,\eta^{(1)}],\\*
    \vdots & \nonumber \\*
    \di\eta^{(n)} &=
    \begin{cases}
      \displaystyle \sum_{k=0}^{n/2 -1} [ \st j, \eta^{(2k+1)} ] + [ j, \eta^{(2k)} ] & \text{if $n$ is even} \\
      [\st j, \eta^{(n-1)}] + \displaystyle \sum_{k=0}^{\left( n - 1
        \right)/2 - 1} [ \st j, \eta^{(2k)} ] + [ j, \eta^{(2k+1)} ] & \text{if $n$ is odd\,.} 
    \end{cases}
  \end{align}
\end{subequations}
Note that $\eta^{(0)}=\text{const.}$ is not the $\eta$ we started
with, but reproduces the $\overline\epsilon$ from the symmetry of the right
action.

Alternatively, the condition $\nabla_J \eta_\zeta=0$ is satisfied if we require
\begin{equation}
  \eta_\zeta = W^{-1}(\zeta)\, \overline \epsilon\, W(\zeta) \, ,
\end{equation}
where $\epsilon \in \mathfrak{g}$ and $W$ is defined by $J=W^{-1} \di
W$ and is given explicitly in Eq.~(\ref{eq:w}).  Since $\eta_{\zeta}$
is a function of $\epsilon$, we can write explicitly this dependence
by introducing the variation operator
\begin{equation}
  \delta( \epsilon, \zeta) g = g\, \eta_{\zeta} = g\, W^{-1} (\zeta)\,
  \overline \epsilon\, W(\zeta) \, . 
\end{equation}
Expanding $\delta(\overline \epsilon; \zeta) $ in series of $\zeta$
\begin{equation}
  \delta(\overline \epsilon; \zeta) = \sum_{n=0}^\infty \delta^{(n)}(\overline \epsilon)\,\zeta^n
  \, , 
\end{equation}
one finds that the $\delta^{(n)}$ form (half of) the \emph{loop algebra} of
the original algebra $\mathfrak{g}$ (see~\cite{Schwarz:1995td}):
\begin{equation}
  \comm{\delta^{(n)} (\overline \epsilon_1)}{\delta^{(m)} (\overline \epsilon_2)} =
  \delta^{(n+m)} (\comm{\overline \epsilon_1}{\overline \epsilon_2}) \, , \hspace{2em}
  \forall n,m = 0, 1, \dots
\end{equation}
The zero modes of the algebra describe the explicit
symmetry $G_R$ and \emph{not} $G_L$ as one could have expected. The algebra
can be extended to a full $\widehat{\mathfrak{g}}$ as shown
in~\cite{Lu:2008kb}. The same procedure
can be repeated for the right current $\overline J(\zeta)$, thus leading to
another infinite set of symmetries commuting with these ones. The full
hidden symmetry algebra is then $\widehat{\mathfrak{g}} \times
\widehat{\mathfrak{g}}$.

\newpage
\section{Deformed hierarchy for the squashed groups}
\label{sec:deform-hier-squash}

In Section~\ref{sec:integr-squash-groups} we found that the Lax
current $\wt J(\zeta)$ can be expanded as
\begin{equation}
  \wt J (\zeta) = \wt J'^{(0)} - \Lambda(\zeta) \wt j \, .  
\end{equation}
The presence of the zero mode changes the hierarchy of variations
described in Appendix~\ref{sec:hier-vari-princ}.

The equations of motion are invariant under the variation
\begin{equation}
  \delta_{\wt \eta_\zeta} g = g\, \wt \eta_\zeta \, , 
\end{equation}
where $\wt \eta_\zeta$ is covariantly constant with respect to
$\wt J'(\zeta)$:
\begin{equation}
  \di \wt \eta_\zeta + \comm{\wt J' (\zeta) }{\wt \eta_\zeta} = 0 \, .
\end{equation}
Writing $\wt J'(\zeta) = \wt J'^{(0)} - \Lambda(\zeta) \wt j$ we see that
\begin{equation}
 \nabla_{\widetilde J'}\widetilde\eta_\zeta= \di \wt \eta_\zeta + \comm{\wt J' (\zeta) }{\wt \eta_\zeta} =
  \nabla_{\wt J'^{(0)}} \wt \eta_{\zeta} - \Lambda(\zeta) \comm{ \wt
    j}{\wt \eta_\zeta} = 0 \, ,
\end{equation}
which can be expanded in powers of $\zeta$ as follows:
\begin{subequations}
  \begin{align}
    \nabla_{\wt J'^{(0)}} \wt \eta^{(0)} & =0\,, \\*
    \nabla_{\wt J'^{(0)}} \wt \eta^{(1)} &= \comm{ \st \wt
      j }{\wt \eta^{(0)}}\,,  \\*
    \nabla_{\wt J'^{(0)}} \wt \eta^{(2)} &= 
    \comm{\st \wt j }{\wt \eta^{(1)}} + \comm{\wt
      j}{\wt \eta^{(0)}}\,, \\*
    \vdots \nonumber \\*
    \nabla_{\wt J'^{(0)}} \wt \eta^{(n)} &=
    \begin{cases}
      \displaystyle \sum_{k=0}^{n/2 -1} \comm{\st \wt j}{\wt
        \eta^{(2k+1)}} + \comm{\wt j}{\wt \eta^{(2k)}} & \text{if $n$
        is even} \\*
      \comm{\st \wt j}{\wt \eta^{(n-1)}} + \displaystyle \sum_{k=0}^{\left(
          n - 1 \right)/2 - 1} \comm{ \st \wt j}{ \wt \eta^{(2k)} } + \comm{ \wt
      j}{ \wt \eta^{(2k+1)} } & \text{if $n$ is odd.}
      \end{cases}
    \end{align}
  \end{subequations}
Observe that the only difference with respect to the hierarchy in the
principal chiral model case obtained in Eq.~(\ref{eq:PCM-hierarchy})
is that the differential has been traded for a covariant derivative
with respect to $\wt J'^{(0)}$. In particular, $\wt \eta^{(0)}$ is
not a constant, but is covariantly constant. One can easily verify
that the variation $\wt \eta(\zeta)$ can be written as
\begin{equation}
  \wt \eta_\zeta= \wt W^{-1}(\zeta) \overline \epsilon \wt W(\zeta) \, , 
\end{equation}
where $\overline \epsilon \in \mathfrak{g}$. At zero order in $\zeta$ this becomes:
\begin{equation}
  \wt \eta^{(0)} (x,t) = ( \wt W^{(0)}(x,t) )^{-1} \overline \epsilon \wt W^{(0)}(x,t) \, . 
\end{equation}

\section{Geodesics on the squashed groups}
\label{sec:geod-squash-groups}

The geodesics of a manifold can be identified with the trajectories of
free point particles. In other words they are the minima of the action
\begin{equation}
  S = \int_\setR \di t \, \left[ G_{ij}(u(t)) \dot u^i(t)
  \dot u^j (t) + \left( \dot z(t) + f_i
    (u(t)) \dot u^i (t)  \right)^2 \right] \, ,  
\end{equation}
which is the one-dimensional counterpart of the action we have
considered in this note.

In the case of a squashed group, this reads
\begin{equation}
  S = \int_\setR \di t \, \left[ - \Tr[ (g^{-1} \dot g)^2 ] + \tanh^2
    \Theta \Tr [ g^{-1} \dot g T_C ]^2 \right] \, ,
\end{equation}
where now $g$ is a map
\begin{equation}
  g : \setR \to G \,,
\end{equation}
The effect of the squashing is to break the left symmetry group $G$
to its Cartan subgroup $T \subset G$.  From the variation
\begin{equation}
  \delta g = \eta g \, , \hspace{2em} \eta : \setR \to \mathfrak{g} \, ,
\end{equation}
we find the conserved right currents
\begin{equation}
  \frac{\di }{\di t } \overline j(t) = \frac{\di }{\di t} \left[ \dot g g^{-1} - \tanh^2 \Theta \, g T_C g^{-1} \Tr [g^{-1} \dot g T_C] \right] = 0\, ,
\end{equation}
while from the variation
\begin{equation}
  \delta g = g \eta \, \hspace{2em} \eta: \setR \to \mathfrak{t} \, ,
\end{equation}
where $\mathfrak{t}$ is the Cartan subalgebra of $\mathfrak{g}$, we
find the conservation of the left currents:
\begin{equation}
  \frac{\di}{\di t} j^\alpha = \frac{\di }{ \di t} \Tr[ T_\alpha g^{-1} \dot g ] = 0 \, , \hspace{2em} \alpha = 1, 2, \dots, \mathop{\mathrm{rank}}(G) \, ,
\end{equation}
where $T_\alpha $ are the generators of $\mathfrak{t}$.

These first order equations can be integrated to give the trajectories
$g = g(t)$ of a geodesic starting from a point $g(0)$. From the left
current conservation we find:
\begin{equation}
 g(t) = g(0) \, \ee^{l_\alpha T_\alpha t} \hspace{2em} l_\alpha \in
  \setR \, , \, T_\alpha \in \mathfrak{t} \, ,  
\end{equation}
while from the right currents:
\begin{equation}
 g(t) = g(0) \, \ee^{R t} \ee^{l_C \tanh^2 \Theta T_C t} \hspace{2em} R
  \in \mathfrak{g} \, , \, l_C \in \setR \, .
\end{equation}
These are to be compared with the usual geodesics on the group manifold
$G$, which are obtained for $\Theta = 0$:
\begin{equation}
  g(t) = g(0) \, e^{L t} \hspace{2em} L \in \mathfrak{g} \, .  
\end{equation}

\bibliography{T-duality,DomenicosPapers,addedrefs}

\end{document}